\documentclass[floatfix,aps,prd,twocolumn]{revtex4}

\usepackage{amsmath}
\usepackage{amsfonts}
\usepackage{amssymb}
\usepackage{bm}
\usepackage{enumerate}
\usepackage[dvips]{color,graphicx}
\usepackage{epsfig}
\usepackage[figuresright]{rotating}

\begin{document}
\preprint{JLAB-THY-09-1096}

\title{Confirmation of quark-hadron duality in the
	neutron $F_2$ structure function}

\author{S.~P.~Malace$^1$, Y.~Kahn$^{2,3}$, W.~Melnitchouk$^3$,
	C.~E.~Keppel$^{3,4}$}
\affiliation{
    $^1$\mbox{University of South Carolina, Columbia,
        South Carolina 29208}               \\
    $^2$Northwestern University, Evanston, Illinois 60208   \\
    $^3$Jefferson Lab, Newport News, Virginia 23606 \\
    $^4$Hampton University, Hampton, Virginia 23668}


\begin{abstract}
We apply a recently developed technique to extract for the first time
the neutron $F_2^n$ structure function from inclusive proton and
deuteron data in the nucleon resonance region, and test the validity
of quark-hadron duality in the neutron.
We establish the accuracy of duality in the low-lying neutron
resonance regions over a range of $Q^2$, and compare with the
corresponding results on the proton and with theoretical expectations.
The confirmation of duality in both the neutron and proton opens the
possibility of using resonance region data to constrain parton
distributions at large $x$.
\end{abstract}

\maketitle

The quest to understand the strong interactions at intermediate
energies, and particularly the transition from quark-gluon to hadron
degrees of freedom, is one of the main outstanding challenges in
modern nuclear physics.
Considerable attention has been focused recently on the 
``duality'' between quark and hadron descriptions of observables
in electron--hadron scattering.
A classic example is the finding \cite{BG} that inclusive structure
functions in the region dominated by the nucleon resonances on average
resemble the structure functions measured in the deep inelastic
scattering (DIS) region at higher energies.

With the availability of high-precision data from Jefferson Lab and
elsewhere, duality has now been firmly established for the proton
$F_2$ and $F_L$ structure functions \cite{IOANA,LIANG,MALACE}, and
exploratory studies in spin-dependent and semi-inclusive scattering
have provided tantalizing glimpses of the flavor and spin dependence
of duality (for a review see Ref.~\cite{MEK}).
A complete picture of the workings of duality in the nucleon can
only be constructed, however, with information on duality in the
{\it neutron}, on which little empirical data exists.

Calculations based on quark models point to intriguing differences
between duality in the proton and neutron \cite{SU6}, and some
arguments even suggest that duality in the proton may be due to
accidental cancellations between quark charges, which do not occur
for the neutron \cite{SJB}.
Confirmation of duality in the neutron would therefore firmly
establish that the phenomenon is not accidental, but rather a
robust feature of nucleon structure functions.
More generally, understanding the transition between the resonance
and DIS regions can lead to better constraints on parton distribution
functions (PDFs) at large momentum fractions $x$, by allowing data
at lower final state hadron masses $W$ to be used in global PDF fits
\cite{ALEKHIN,CTEQX}.
Precise knowledge of large-$x$ PDFs, which are currently poorly
constrained, is vital in searches for new physics beyond the Standard
Model \cite{KUHLMANN}, for instance, as well as in neutrino oscillation
experiments \cite{OSC}.

In this Letter we use a recently introduced technique \cite{KMK} to
extract for the first time the neutron $F_2^n$ structure function from
proton ($p$) and deuteron ($d$) $F_2$ data in the resonance region
over a range of photon virtualities from $Q^2 = 0.6$ to 6.4~GeV$^2$.
The new method is based on an iterative approach in which the nuclear
corrections are applied additively, and has been found to accurately
reproduce neutron structure functions of almost arbitrary shape in
both the DIS and resonance regions \cite{KMK}.

The extraction of reliable neutron information from deuterium data
requires a careful treatment of nuclear effects \cite{NP}, and we use
the latest theoretical developments which allow the deuteron structure
function to be analyzed in both the resonance and DIS regions, at both
low and high $Q^2$.
%
%
In the weak binding approximation the deuteron $F_2^d$ structure
function can be written as a sum of smeared proton and neutron
structure functions $\widetilde{F}_2^N$ ($N=p, n$), and an additive
term which accounts for possible modification of the structure
functions off-shell \cite{KMK,KP,AKL},
\begin{eqnarray}
F_2^d &=& \widetilde{F}_2^p + \widetilde{F}_2^n
       +  \delta^{(\rm off)} F_2^d\, .
\label{eq:F2d_conv}
\end{eqnarray}
The smeared nucleon structure functions are given by convolutions
of the nucleon light-cone momentum distribution in the deuteron,
$f_{N/d}$, and the bound nucleon structure functions \cite{KMK,KP},
\begin{eqnarray}
\widetilde{F}_2^N &=& f_{N/d} \otimes F_2^N\, ,
\label{eq:smear}
\end{eqnarray}
where the symbol $\otimes$ denotes a convolution.
The nucleon momentum distribution (or smearing) function $f_{N/d}$ 
accounts for the effects of the nucleon's Fermi motion and binding, 
including finite-$Q^2$ corrections \cite{KMK,KP}, and is taken to be 
identical for the proton and neutron.
The off-shell correction $\delta^{(\rm off)} F_2^d$ has been found in
several models \cite{KP,AKL,MST} to be typically of the order 1--2\%
for $x \lesssim 0.9$.

%
To account for the quasi-elastic (QE) tail in the deuteron data the
elastic nucleon contribution is smeared using the same $f_{N/d}$.
Subtracting from the deuteron $F_2^d$ the QE contribution,
together with the off-shell correction and the smeared proton
$\widetilde{F}_2^p$, one obtains an effective smeared neutron
structure function $\widetilde{F}_2^n$ and then solves
Eq.~(\ref{eq:smear}) for the neutron.

The nuclear effects are parametrized by an additive correction \cite{KMK},
\begin{eqnarray}
\widetilde{F}_2^n
&=& {\cal N} F_2^n + \delta f \otimes F_2^n\, ,
\end{eqnarray}
where $\delta f$ gives the finite width of the smearing function
and ${\cal N}$ its normalization.
The $F_2^n$ structure function is then extracted using an iterative
procedure \cite{KMK}, which after one iteration gives
\begin{equation}
F_2^{n(1)}
= F_2^{n(0)}
+ \frac{1}{\cal N}
  \left( \widetilde{F}_2^n - f \otimes F_2^{n(0)} \right)\, ,
\label{eq:add}
\end{equation}
starting from a first estimate $F_2^{n(0)}$, and iterated until
convergence is reached.
The robustness of this method and its ability to reliably estimate
errors on the extracted neutron function were investigated
extensively for smooth functions in Ref.~\cite{KMK}.
Since the smearing function is sharply peaked, the convergence
of this method is typically extremely fast, requiring only one or
two iterations before the $F_2^d$ function reconstructed from the
extracted $F_2^n$ matches the original data to within experimental
uncertainties.

\begin{figure}
\includegraphics[width=8.5cm]{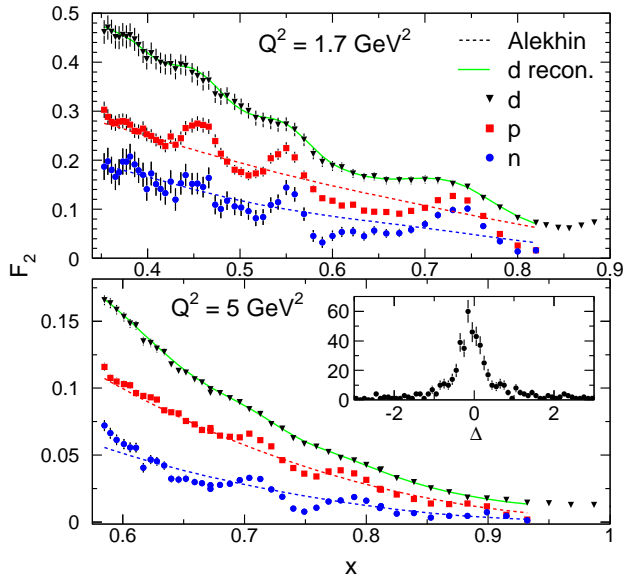}
\caption{
	Extracted neutron $F_2^n$ structure function at $Q^2 = 1.7$
	and 5~GeV$^2$, together with proton and deuteron data, and
	the reconstructed deuteron (total uncertainties are systematic
	and statistical errors added in quadrature).
	The proton and neutron data are compared with the global QCD
	fit from Alekhin {\it et al.} \cite{ALEKHIN}.  The dependence
	of the iteration on the initial value is illustrated in the
	insert (see text).}
\label{fig:F2}
\end{figure}

In this analysis we use proton and deuteron $F_2$ data from JLab
experiment E00-116 \cite{MALACE} and SLAC experiments E49a6 and
E49a10 \cite{SLAC_data}.
The former span the high-$Q^2$ region, $4.5 \leq Q^2 \leq 6.4$~GeV$^2$,
while the latter cover the lower $Q^2$ range,
$0.6 \leq Q^2 \leq 2.4$~GeV$^2$, providing a total of 514 data points.
Because the extraction method requires proton and deuteron $F_2$ data
at fixed $Q^2$, the centering of the data at the same $Q^2$ was made
at the cross section level using the $p$ and $d$ fits from
Ref.~\cite{CHRISTY_FIT}.
To test the sensitivity of the results to the choice of the
bin-centering fit, an additional fit was used for each target 
\cite{MALACE,KEPPEL,GOMEZ} and half the difference in the results
assigned as a systematic uncertainty.

The stability of the iteration method relies on the availability of
relatively smooth data, especially for deuterium (irregularities in
proton data are smoothed out by the smearing).
This is critical at large $x$ where the structure functions are small,
and discontinuities could even render the extracted neutron results
negative.
It is particularly important that the QE contribution to the deuteron
$F_2^d$ be accounted for in the analysis, and we model this using
the same smearing function, $f_{N/d}$, and nucleon form factors from
Refs.~\cite{AMT,BOSTED}.
This is found to provide a good description of the QE peak as a
function of $Q^2$.

An example of the extracted neutron $F_2^n$ structure function is
displayed in Fig.~\ref{fig:F2} for $Q^2$ = 1.7 and 5~GeV$^2$, together
with the input proton and deuteron data (the complete data set will
be shown in Ref.~\cite{NEXT}).
The starting value of the neutron for the iteration was
$F_2^{n(0)}=F_2^p$, and the deuteron $F_2^d$ reconstructed from the
proton and extracted neutron was found to be in good agreement with
the data after two iterations.
The spectrum of the $F_2^n$ structure function in the resonance region
displays similar characteristics as observed from the proton spectrum:
one finds three resonant enhancements which fall with $Q^2$ at
a similar rate as for the proton.

To check that the extracted neutron structure function does not depend
on the starting value of the iteration, the extraction procedure was
repeated assuming a different boundary condition, $F_2^{n(0)}=F_2^p/2$.
The difference between the two results 
$\Delta = [F_2^n(F_2^{n(0)}=F_2^p)
	-  F_2^n(F_2^{n(0)}=F_2^p/2)]$/$\sigma(F_2^n)$,
normalized by the total $F_2^n$ uncertainty $\sigma(F_2^n)$,
is shown in the insert of Fig.~\ref{fig:F2} after two iterations.
One finds an almost Gaussian distribution centered around 0
(the mean of the distribution is around $-0.07$) with a width
well within the typical total uncertainty of $F_2^n$.
In fact only 6\% of the total number of data points lie outside
of a 2$\sigma$ range.
More extreme boundary conditions, such as $F_2^{n(0)}=0$, do not
alter the characteristics of the extracted $F_2^n$ structure function
spectrum, with the resonant structures already visible after just 1
iteration.
On the other hand, as discussed in Ref.~\cite{KMK}, more iterations
are needed for poor choices of initial values, which increases the
scatter of data points if the deuterium data in particular display
any nonuniformities.

The effect of the off-shell correction $\delta^{(\rm off)} F_2^d$
was taken into account using the model of Ref.~\cite{MST}, which
gives $\approx -1.5\%$ correction over most of the $x$ range
considered, and was argued to provide an upper limit on the
correction.
The $F_2^d$ data are corrected by subtracting half of the off-shell
correction from \cite{MST} and assigning a 100\% uncertainty.
When propagated into the $F_2^n$ uncertainty this was found
to contribute less than 2\% to the total error.

In Fig.~\ref{fig:F2} we also show $F_2^p$ and $F_2^n$ from global QCD
fits to DIS data (with $W^2 > 4$~GeV$^2$) from Alekhin {\it et al.}
\cite{ALEKHIN}, which illustrates the striking similarity between the
QCD fit and the resonance data, reminiscent of Bloom-Gilman duality
\cite{BG}.
To quantify this duality we consider ratios of ``truncated'' moments
$M_2$ \cite{TRUNC},
\begin{equation}
M_2(Q^2,\Delta x)
= \int_{\Delta x} dx\, F_2(x,Q^2)\, ,
\end{equation}
in the resonance region for specific intervals $\Delta x$.
Following previous proton data analyses \cite{IOANA,MALACE},
we consider the regions
\begin{itemize}
\item{$1^{\rm st}$ resonance region $\to$ $W^2$ $\in$ [1.3, 1.9] GeV$^2$ }
\item{$2^{\rm nd}$ resonance region $\to$ $W^2$ $\in$ [1.9, 2.5] GeV$^2$ }
\item{$3^{\rm rd}$ resonance region $\to$ $W^2$ $\in$ [2.5, 3.1] GeV$^2$ }
\end{itemize}
as well as the entire resonance region $1.3 \leq W^2 \leq 4$~GeV$^2$.
At a given $Q^2$, the lowest-$W$ ($\Delta$ resonance) region corresponds
to the highest-$x$ range, and for a fixed $W$ interval the larger the
$Q^2$, the higher the~$x$.

\begin{figure}
\includegraphics[width=8.5cm]{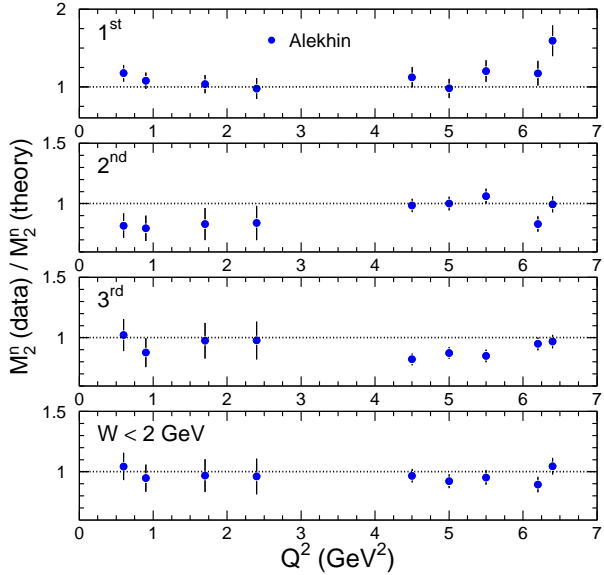}
\caption{
        Truncated neutron moments $M_2^n$ (``data'') in various
	resonance regions
	($1^{\rm st}$, $2^{\rm nd}$, $3^{\rm rd}$ and $W<2$~GeV)
	relative to the QCD fit from Alekhin {\it et al.}
	\cite{ALEKHIN} (``theory'').}
\label{fig:mom}
\end{figure}

The ratio of the truncated moments of the resonance data to the
global QCD fit \cite{ALEKHIN}, computed over the same $x$ range,
is shown in Fig.~\ref{fig:mom} as a function of $Q^2$.
Globally, the agreement between the QCD fit and the resonance data
is quite remarkable, with deviations of $\lesssim 10\%$ observed over
the entire $Q^2$ range.
Locally, in the individual resonance regions the deviations are
generally $\lesssim 15-20\%$, somewhat larger only in the $1^{\rm st}$
resonance region at the largest $Q^2$.
This is not surprising given the fact that the $\Delta$
region at $Q^2 = 6.4$~GeV$^2$ covers the highest-$x$ regime studied,
$x \sim 0.9$, where the QCD fit is mostly beyond its limit of
applicability.

\begin{figure}
\includegraphics[width=8.5cm]{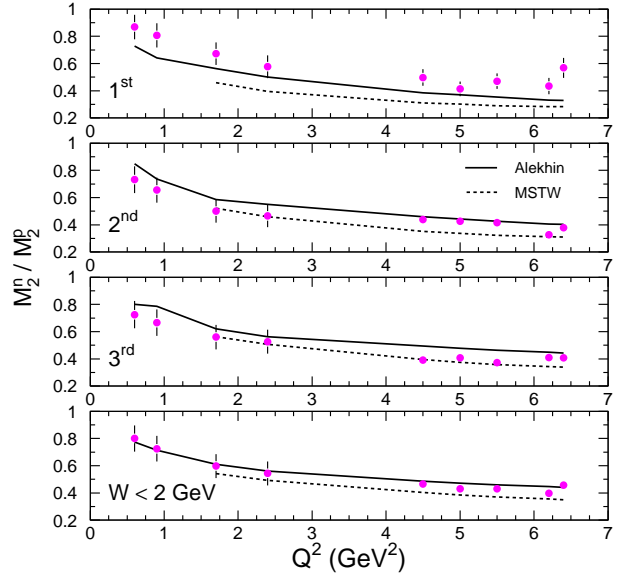}
\caption{
        Ratio of truncated neutron to proton moments $M_2^n/M_2^p$
	in various resonance regions as a function of $Q^2$, compared
	with global QCD fits from Alekhin {\it et al.} \cite{ALEKHIN}
	and MSTW \cite{MSTW}.}
\label{fig:mom_np}
\end{figure}

The isospin dependence of duality can be studied by comparing the
truncated neutron moments with the analogous proton moments.
The ratio of these is displayed in Fig.~\ref{fig:mom_np} as a
function of $Q^2$ for the various resonance regions, and compared
with global QCD fits from Alekhin {\it et al.} \cite{ALEKHIN} and
from MSTW \cite{MSTW}, corrected for target mass effects \cite{TMC}.
The MSTW fits are shown for $Q^2 \gtrsim 2$~GeV$^2$, which corresponds
to their approximate limit of validity.

The ratios show good agreement with the data, with the exception
of the $\Delta$ region which is somewhat underestimated.
Since the proton and neutron transitions to the $\Delta$ are isovector,
the resonant contributions should be identical; on the other hand,
the DIS structure functions in the $\Delta$ region are expected to
be rather different, with $F_2^n \ll F_2^p$, so that violation of
duality here is expected to be strongest.
In addition, the QCD fits are least constrained in this region
due to the scarcity of large-$x$ DIS data.
This is especially the case for the MSTW fit \cite{MSTW} which
limits the data sets to $W^2 > 15$~GeV$^2$.

\begin{figure}
\includegraphics[width=8.5cm]{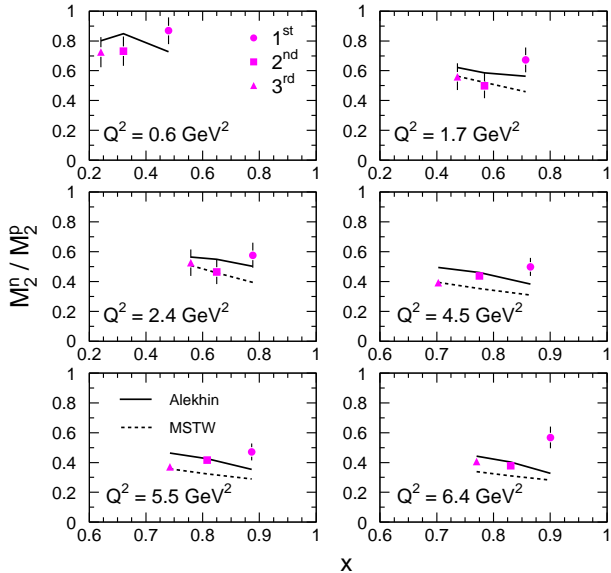}
\caption{
        As in Fig.~\ref{fig:mom_np} but as a function of $x$
	for fixed $Q^2$.}
\label{fig:mom_x}
\end{figure}

The $M_2^n/M_2^p$ ratios at fixed $Q^2$ are shown in
Fig.~\ref{fig:mom_x} as a function of $x$ for the three resonance
regions, compared with the QCD fits as in Fig.~\ref{fig:mom_np}.
The global fits offer a good description of the $2^{\rm nd}$ and
$3^{\rm rd}$ resonance region data, revealing clear evidence of
duality down to $Q^2$ as low as 0.6~GeV$^2$.
The fits underestimate the $\Delta$-region ratios 
and this trend becomes more pronounced as one moves to larger $Q^2$
($\gtrsim 4$~GeV$^2$) and larger $x$.
The Alekhin {\it et al.} fit \cite{ALEKHIN} offers a better
description at large $x$, which is likely due to its
inclusion of lower-$W$, lower-$Q^2$ data.

Our results can be compared with quark model expectations for the
isospin dependence of duality, which predict systematic deviations
of resonance data from local duality.
Assuming dominance of magnetic coupling, the proton data are expected
to overestimate the DIS function in the $2^{\rm nd}$ and $3^{\rm rd}$
resonance regions due to the relative strengths of couplings to
odd-parity resonances, especially those in the quark spin-$\frac12$
octet \cite{SU6}.
The neutron data are predicted to lie below the DIS curve in the
$2^{\rm nd}$ resonance region due to the small coupling to octet
states with spin $\frac12$, but have larger couplings to odd-parity
quark spin-$\frac32$ octet states.
Remarkably, the neutron data do indeed underestimate the global
$F_2^n$ fits in the $2^{\rm nd}$ resonance region, just as the proton
data were found to exceed the global $F_2^p$ fits \cite{IOANA,MALACE}.
Moreover, the similarity between the truncated $M_2^n$ moments in the 
$W^2 < 4$~GeV$^2$ and DIS regions strongly suggests that the resonance
cancellations in the proton are not accidental \cite{SJB}, but rather
form a systematic pattern which dramatically reveals itself through
the Bloom-Gilman duality phenomenon.

In conclusion, we have extracted the neutron structure function $F_2^n$
for the first time in the resonance region from inclusive proton and
deuteron data.
Our comparisons of empirical truncated moments to those extracted
from global QCD fits to high-$W^2$ data show clear signatures of
Bloom-Gilman duality, with better than $15-20\%$ agreement in the
2$^{\rm nd}$ and 3$^{\rm rd}$ resonance regions, and less than 10\%
deviations when integrated over the entire $W^{2}<4$~GeV$^{2}$ region.
The confirmation of duality in the neutron establishes that the
phenomenon is not accidental, but is a general property of nucleon
structure functions.
Our findings suggest that averaged resonance data could in future
be used to constrain the large-$x$ behavior of global QCD fits 
\cite{ALEKHIN,CTEQX} by relaxing the $W^{2}$ cuts on data down
to the 2$^{\rm nd}$ resonance region, $W^2 \sim 1.9$~GeV$^2$.
This could also have significant impact for searches for new physics
beyond the Standard Model at colliders and neutrino oscillations
experiments.

\begin{acknowledgments}

We thank S.~Alekhin, S.~Kulagin and G.~Watt for helpful communications.
This work was supported by the DOE contract No. DE-AC05-06OR23177,
under which Jefferson Science Associates, LLC operates Jefferson Lab,
and by NSF grant PHY-0856010.

\end{acknowledgments}


\end{document}